\documentclass[aps,prl,twocolumn,floatfix,amsmath,amssymb,superscriptaddress,bibnotes,nolongbibliography,footinbib,fontsize=10]{revtex4-2}
\usepackage[T1]{fontenc}
\usepackage[latin1]{inputenc}
\usepackage[english]{babel}

\usepackage{setspace}

\usepackage{amsmath}
\usepackage{amssymb}
\usepackage{amsfonts}
\usepackage{mathtools}	

\usepackage[output-decimal-marker={.}, separate-uncertainty]{siunitx}

\usepackage{booktabs}
\usepackage{caption}
\usepackage{ragged2e}
\usepackage{graphicx}
\usepackage{array}
\usepackage{xcolor}
\usepackage{colortbl}
\usepackage{subfig}
\usepackage{mathrsfs}
\usepackage{wrapfig}

\usepackage{hyperref}
\hypersetup{
colorlinks=true,final=true,
linkcolor=blue,
citecolor=blue,
filecolor=blue,
urlcolor=blue,
}

\usepackage{lipsum}
\usepackage{url}
\usepackage{appendix}

\usepackage{notes2bib}

\begin{document}


\title{Stochastic resonance in disordered charge-density-wave systems}

\author{Francesco Valiera}
\affiliation{Institut f{\"u}r Theoretische Physik, Universit{\"a}t Hamburg, Notkestra{\ss}e   9 , 22607 Hamburg, Germany}
\affiliation{The Hamburg Centre for Ultrafast Imaging, Luruper Chaussee 149, 22761 Hamburg, Germany}

\author{Antonio Picano}
\affiliation{JEIP, UAR 3573, CNRS, Coll{\`e}ge de France, PSL Research University,
11 Place Marcelin Berthelot, 75321 Paris Cedex 5, France}

\author{Martin Eckstein}
\affiliation{Institut f{\"u}r Theoretische Physik, Universit{\"a}t Hamburg, Notkestra{\ss}e   9 , 22607 Hamburg, Germany}
\affiliation{The Hamburg Centre for Ultrafast Imaging, Luruper Chaussee 149, 22761 Hamburg, Germany}

\begin{abstract}
Ultrafast disordering observed after photo-excitation challenges the conventional picture of photo-induced transitions where symmetry-breaking takes place along a single collective coordinate. We propose that key spectroscopic signatures of these transient disordered states can be revealed through stochastic resonance, a hallmark of nonlinear stochastic dynamics. Studying the disordered phase of Holstein model we show that, at given frequency, the linear response as a function of temperature has a peak, which indicates enhanced coherent switching between metastable configurations. From this resonance, we extract the intrinsic stochastic transition timescale and energy barrier separating equivalent local minima. This mechanism offers a new perspective to identify and characterize hidden disordered phases in driven many-body systems.
\end{abstract}

\maketitle
When studying the ultra-fast light-induced dynamics of quantum materials~\cite{delatorre_kennes:ultrafast_control_review,murakami:photoinduced_noneq_mott_insulators_review}, the role played by disorder and inhomogeneities appears to be significant in many cases, but is  far from understood. Particularly intriguing is the spontaneous emergence of disorder in nominally clean systems, which can happen due to preformed order close to a symmetry-breaking phase transition. Microscopic degrees of freedom then locally reflect the symmetry-broken state, but no global order is established.  This mechanism was proposed for structural phase transitions, where ions in each unit cell $j$ would undergo a finite displacement $X_j$, but randomly select one of the symmetry-equivalent positions of the ordered phase~\cite{wall:disorder_C4-nature_comm,wall:ultrafast_disordering_VO2-science,wall:ultrafast_disordering_collisional_forces-nature_physics,wall:ultrafast_xray_V02-nature_physics,wall:prl}.  The average displacement $\overline{X}$ then remains zero and therefore a (time-dependent) Ginzburg--Landau theory for $\overline{X}$ fails to capture the system's local state or its dynamics.
 
A direct probe of (dynamically emerging) disorder in correlated systems is diffuse scattering, which can be used  in both in equilibrium~\cite{osborn_pelc:diffuse_scattering_correlated_electron_systems-science_adv} and in pump-probe experiments~\cite{wall:prl,wall:ultrafast_disordering_collisional_forces-nature_physics,wall:ultrafast_disordering_VO2-science,wall:ultrafast_xray_V02-nature_physics}. Moreover, time-resolved microscopy has also been 
used to characterize materials at the microscopic level in a dynamical setting~\cite{danz-domroese-ropers:ultrafast-nanimaging-order-parameter}. 
Probing directly the evolution of inhomogeneous order parameters on ultrafast timescales, however, remains challenging, and complementary spectroscopic methods would be desirable.  A dynamical signature of the preformed local order could be given by the transitions between the different local configurations. At finite temperature, however, the latter 
are stochastic rather than coherent, and should therefore not give rise to a clear resonance in the frequency-dependent response.
In this work we propose that  in this noise-dominated regime the existence of degenerate local configurations can nevertheless  be identified with the help of stochastic resonance \cite{stochastic_resonance_review,jung_hanggi:stoc_res_PRA}. Stochastic resonance,  a phenomenon originally introduced in the context of climate dynamics~\cite{benzi:stochastic_resonance,benzi:stochastic_resonance2}, arises because in a nonlinear dynamical system such as a particle in a bistable potential, noise can not only reduce, but also amplify the response to a periodic perturbation. 
The amplification is maximal when the noise-induced transition rate between states becomes commensurate with the driving frequency  -- hence the term resonance. 
Through the stochastic resonance, the relevant timescales of a purely stochastic process without any coherent oscillation could therefore potentially be detected in a coherent spectroscopic experiment, and then be used to characterize the disordered potential energy landscape of the system. Stochastic resonance occurs across a wide range of contexts, from climate dynamics to biology and neuroscience~\cite{stochastic_resonance_review, stochastic_resonance_review_new}. 
Within condensed matter  and many-body physics,
the phenomenon  has been discussed for spin chains and qubit-based models~\cite{stoc_res_quantum_many_body,stoc_res_jaynes_cummings},  cold atomic and Rydberg systems~\cite{stoc_res_cold_atoms,quantum_stoc_res_Ryd}, large molecular systems~\cite{stoc_res_polaron_ratchets}, and semiconductor and graphene-based devices~\cite{stoc_res_GaAs, stoc_res_graphene}. Motivated by this, the present work aims to relate stochastic resonance to the response of inhomogeneous phases originating from preformed symmetry-broken states in condensed matter.

{\em Model and method -- } 
Specifically, we consider the Holstein model,
which captures the essential physics of electrons interacting with lattice vibrations~\cite{holstein_original1,holstein_original2}. The Hamiltonian reads
\begin{equation}
\label{eq:holstein}
H = -J_0\sum_{\langle i, j\rangle,\sigma}c^\dagger_{i\sigma}c_{j\sigma} + \sqrt{2}g\sum_j X_j(n_j - 1) + H_\mathrm{ph},
\end{equation}
where $J_0$ is the electron hopping amplitude, and $c_{j\sigma}$ ($c^\dag_{j\sigma}$) is the annihilation (creation) operators for an electron in the $j$-th unit cell. The local electron density $n_j = c^\dag_{j\uparrow}c_{j\uparrow} + c^\dag_{j\downarrow}c_{j\downarrow}$ is coupled to the displacement $X_j$ of an ion in the same unit cell, with interaction constant $g$. The phonon Hamiltonian is given by $H_\mathrm{ph} = \sum_j (\Omega/2)(P_j^2 + X_j^2)$. 
The Holstein model on a bipartite lattice provides a minimal description of a charge density wave (CDW) formation: below a critical temperature $T_c$, the sublattice symmetry is broken, and both the ion displacement $\langle X_j\rangle$ and the electron density $\langle n_j\rangle$ change sign between the sublattices $A$ and $B$. A suitable order parameter is the difference $\mathcal{O} = \langle X \rangle_A - \langle X \rangle_B$ of the mean ion displacements on the two sublattices.

To study the dynamics of emergent disordered states, we will use a semiclassical stochastic approximation (SC)~\cite{Kamenev, picano:semiclassical_theory,dutta_majumdar:noneq_thermal_state_Mott_insulator}. Within this approach, the displacements $X_j$ are treated as classical time-dependent observables, while the electrons retain their quantum character. Each ion then obeys the stochastic differential equation \cite{Gardiner}
\begin{equation}
\label{eq:stoc_diff_eq}
\ddot{X}_j = -\Omega^2 X_j + F_\mathrm{el}\,(X) - \Omega 
\Gamma_j\,(X)
\dot{X}_j + \Omega 
\sqrt{K_j\,}\xi_j\,(t),
\end{equation}
where the anharmonic force, the damping  and the noise amplitude $K$ all derive from the electron-phonon coupling. 
In particular, $F_\mathrm{el}$ is the Ehrenfest force generated by the electrons on the ions,
\begin{equation}
\label{eq:electronic_force}
F_\mathrm{el}\,(X) = \sqrt{2}g\Omega \left(\langle n_j(t)\rangle_X-1\right),
\end{equation}
where the effect of the ions on the electrons is made clear by the $X$-dependence of the expectation value $\langle n_j(t)\rangle_X$
($X\equiv\{X_j\}$ denotes the entire displacement configuration); 
$\xi_j\,(t)$ represents white noise, i.e., a 
Gau\ss{}ian 
stochastic process with zero mean and autocorrelation $\langle\!\langle \xi\,(t)\xi\,(t')\rangle\!\rangle=\delta\,(t-t')$~\cite{Gardiner}, where $\langle\!\langle\cdots\rangle\!\rangle$ is the average over the noise realizations.
The damping $\Gamma_j$ and the 
noise amplitude $K_j$ are both related to the electronic density-density response function (see~\cite{picano:inhomogeneous_disordering,picano:semiclassical_theory} and the Supplemental Material~\cite{suppl}{}).

Due to the locality of the interaction in~\eqref{eq:holstein}, the electronic state of the model can be described within DMFT~\cite{georges-rev,aoki}. Each sublattice is mapped to an effective impurity model with  one oscillator and one electronic orbital coupled to an electronic bath with a self-consistent hybridisation function. To describe generic qualitative results, 
we will  focus on the Bethe lattice in the limit of infinite coordination number $Z$, which gives rise to the self-consistency condition
\begin{equation}
\label{eq:dmft1}
\Delta_{o}\,(\omega,t) = \frac{1}{Z} \sum_{j} J_{oj}^2 G_{j}\,(\omega,t) 
\end{equation}
where $\Delta_{j}$ and  $G_{j}$ are respectively the bath hybridisation function and the impurity Green's function at a site $j$, and the sum extends  over the nearest neighbors of $o$.  In the semiclassical approach for $Z\to\infty$, this sum turns into a statistical average 
$\langle\!\langle G_j\rangle\!\rangle$ on the opposite sublattice (similar to DMFT for disordered systems~\cite{miranda_dobrosavljevic:dmft_correlation_disorder}),
\begin{equation}
\label{eq:dmft2}
\Delta_{A/B}\,(\omega,t)=
J_0^2\,\langle\!\langle G_j(\omega,t)\rangle\!\rangle.
\end{equation}
Through this self-consistency condition the hybridization functions $ \Delta_{A/B}$ implicitly depend on all configurations $X$ at a given time, leading to non-trivial coupled dynamics between the subsystems. 
Taking $J_0$ as unit of energy, the free density of states has a semielliptic shape with bandwidth $W=4$; moreover, the bare phonon frequency was taken as $\Omega=0.2$.
For the results below, an ensemble of $N_{\rm ens}=250$ trajectories for each sublattice was simulated.

For the simulation, two further approximations were made (for more details, see also the Supplemental material~\cite{suppl}{}): firstly, the electronic state is computed within the common adiabatic approximation at a given temperature $T$.  In Ref.~\cite{picano:inhomogeneous_disordering}, the electronic state was propagated explicitly using a quantum Boltzmann equation~\cite{Picano_Li:QBE}. Nonetheless, 
in the present case the differences with the adiabatic approach are found to be small, because 
electrons couple to other degrees of freedom that act as a thermostat, as verified below. 
Secondly, 
we take the damping coefficient $\Gamma$, which, in principle, could be obtained from the density--density correlation function at fixed $ X $~\cite{picano:inhomogeneous_disordering}, to be independent of $ X $.
This is justified because we focus on the disordered phase, where the electronic system remains ungapped, such that the density--density correlation function depends only weakly on $ X $.  The adiabatic approximation then implies the Einstein relation 
$ K = 2\Gamma / \beta $ for the diffusion coefficient.

\begin{figure*}
\includegraphics[width=\textwidth]{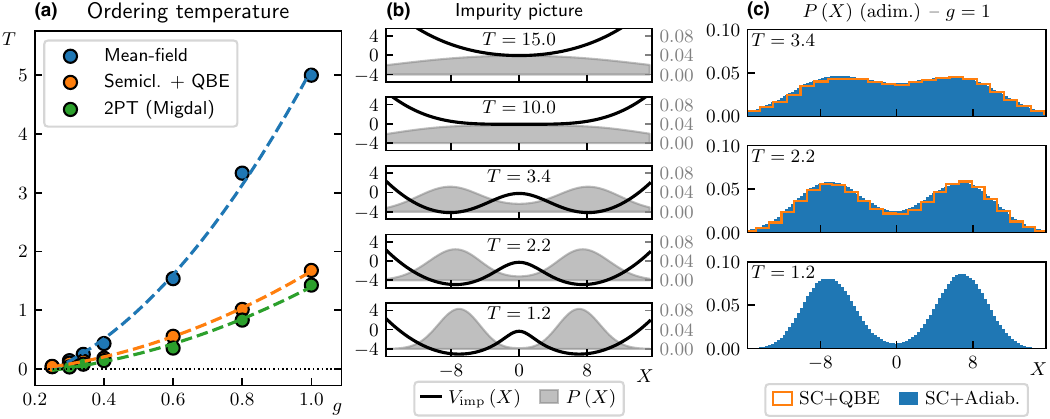}
\caption{\label{fig1}  \justifying  (a) Ordering temperature of the Holstein model calculated with various methods.
(b) Mean-field potential for one ion in the impurity model (black curve, left y axis) and corresponding Boltzmann distribution (grey area, right y axis). The energy scale is set by choosing electronic density of states (hybridization function) with a semi-circular shape and width $W=40$; the hybridization strength is then chosen to match the fully interacting local density of states of the lattice simulations. The coupling constant was set to $g=1$ and the phonon frequency to $\Omega=0.2$. (c) Probability distributions for the ions in the Bethe lattice at $g=1$ calculated both with QBE and adiabatic approach.
}
\end{figure*}

{\em Results -- } 
The semiclassical framework has already been used to capture the photoinduced formation of an inhomogeneous state with local 
$\mathbb{Z}_2$-symmetry breaking \cite{picano:inhomogeneous_disordering}: after long-range order is melted by a laser pulse, each ion effectively moves in a bistable potential, randomly selecting one of two symmetry-equivalent minima. 
Moreover, classical fluctuations enable rare stochastic transitions between the minima.
The emergent bistability can already be understood at the level of a single Holstein impurity embedded in a metallic host, that is, a system where the oscillator is present only at one site $j$. Within a mean field description, the displacement $X_j$ experiences an effective potential $V_\textup{imp}\,(X_j)$, which is the sum of the harmonic $X_j^2$ potential and the free energy of the electrons at given $X_j$. Below a critical temperature $T_{\rm c,imp}$, the potential becomes bistable and the impurity undergoes a symmetry-breaking transition, as shown in Figure~\ref{fig1}b (see the caption for the choice of the model parameters). For the calculation of the potential, we refer to the Supplemental Material~\cite{suppl}.

The symmetry breaking in the impurity model is a mean-field artifact, but it captures key aspects of the exact solution, where the ion occupies the two minima of the potential with equal probabilities, resulting in a bimodal distribution of displacements $P(X_j)$. As found in \cite{picano:inhomogeneous_disordering}, the semiclassical stochastic approximation well reproduces the exact distribution function $P(X_j)$ obtained from Quantum Monte Carlo (QMC)~\cite{werner_millis:dmft_holstein_hubbard} in an intermediate temperature regime where quantum tunneling is not 
dominant.  Given  this agreement, we will extend the semiclassical simulation to evaluate the real-frequency response functions, which are difficult to obtain from QMC, and look for the signatures of stochastic resonance. We will investigate this question directly for the full lattice model rather than the impurity model.

In Figure~\ref{fig1}a the ordering temperature is shown as a function of the coupling constant. Results from the semiclassical stochastic DMFT approach are contrasted with mean-field theory and the second-order perturbation theory (Migdal approximation) 
\cite{murakami_werner_tsuji_aoki:interaction_quench_holstein,randi}. 
Mean-field theory predicts the highest critical temperature $T_c^{\rm mf}$, of order $g^2/\Omega$, which is lowered by including fluctuations within the other two approximations.  The agreement of the latter two approximations provides further validation of the semiclassical approach in that parameter regime.

To identify inhomogeneously disordered phases, one can examine the distribution $P_{A/B}(X)$ of the ion displacements $X$ on each sublattice. For $T > T_c$, sublattice symmetry is unbroken and $P_A(X) = P_B(X) \equiv P(X)$. Over a broad temperature range above $T_c$, this distribution exhibits a double-peak structure, indicating that each ion effectively moves in a double-well potential (two upper panels of Fig.~\ref{fig1}c). Below $T_c$, $P_{A/B}(X)$ would evolve into a single peak centered around $X_{A/B} = \pm \mathcal{O}$, reflecting the emergence of long-range order.
In equilibrium, we find that the bistability extends roughly from $T_c$ up to the mean-field critical temperature $T_c^{\rm mf}$. A bistable regime without global symmetry breaking can transiently appear after laser excitation below $T_c$,  as proposed in the context of ultrafast disordering~\cite{wall:ultrafast_disordering_VO2-science,wall:disorder_C4-nature_comm,wall:ultrafast_disordering_collisional_forces-nature_physics,wall:ultrafast_xray_V02-nature_physics,wall:prl}.
Such a phase was also found in the Holstein model~\cite{picano:inhomogeneous_disordering}
In the present simulation, we can simply prepare a disordered state with global sublattice symmetry below $T_c$ by symmetrizing the self-consistent hybridization function \eqref{eq:dmft2}, see lower panel of Fig.~\ref{fig1}c.

\begin{figure*}
\includegraphics[width=\textwidth]{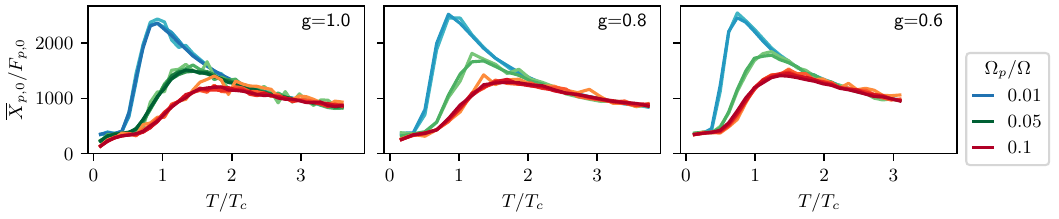}
\caption{\label{fig2} \justifying Linear response of the ions under the effect of a periodic perturbation, shown as a function of temperature. Each panel corresponds to a different coupling constant and each set of curves to a different driving frequency (see the legend on the right).
For fixed driving frequency, an increasing color intensity of the curves correspond to an increase of the maximum driving force $F_{p,0}$.
In particular, the values used were, in units of $J_0^{5/2}\Omega^{-1/2}$:  
$2.22\times10^{-3}$, $4.44\times10^{-3}$, $6.66\times10^{-3}$, $8.88\times10^{-3}$ and $11.1\times10^{-3}$.
}
\end{figure*}

Bistability and noise are basic requirements for the realization of stochastic resonance. In order to witness this phenomenon, we add a periodic force $F_p\,(t)= F_{p,0} \cos{(\Omega_p t)}$ with amplitude $F_{p,0}$  and driving frequency $\Omega_p$ to Eq.~\eqref{eq:stoc_diff_eq}; the latter will be much smaller than the characteristic frequency of oscillation inside each potential well ($\Omega_p\ll \Omega$). 
To assess the stochastic resonance, we  measure the ensemble average $\overline{X}\,(t) = \langle\!\langle X_j\rangle\!\rangle$~\cite{jung_hanggi:stoc_res_PRA,stochastic_resonance_review}.
The amplitude $F_{p,0}$ is taken small enough such that the system remains in the 
linear regime, where  $\overline{X}\,(t)$ can be written as
\begin{equation}
\begin{split}
\overline{X}\,(t) 	
					&= \overline{X}_{p,0}\cos{(\Omega_p t+\phi_0)} + \Delta\overline{X}\,(t)\,.
\end{split}
\end{equation}
The first term is the linear response, and the second term is a fluctuating contribution which will vanish for large statistical ensembles. The linear response $\overline{X}_{p,0}/F_{p,0}$ is therefore extracted from the Fourier transform of the average position $\overline{X}_{p,0} = \frac{2}{t_\textup{max}}\int_0^{t_\textup{max}}\mathrm{d}{t}\,\mathrm{e}^{-i\omega t} \langle\!\langle X_j (t)\rangle\!\rangle$ at the driving frequency $\omega=\Omega_p$. The simulation time $t_{\rm max}$ was chosen in all cases around $12\, \times \,2\pi /\Omega_p$.

Stochastic resonance becomes apparent if the response is analysed at fixed $\Omega_p$ as a function of the noise intensity, which is quantified by temperature. The results  are shown in Figure~\ref{fig2}, where each panel corresponds to a specific coupling constant, with traces for three values $\Omega_p/\Omega=0.1,0.05,0.01$.  The peak structure of the curves in Figure~\ref{fig2} is the typical signature of stochastic resonance, indicating an optimal noise intensity which maximizes the coherent response.  For stochastic resonance, the  optimal temperature $T_{\rm sr}(\Omega_p)$ is  determined by matching one half period $\pi/\Omega_p$ of the driving frequency with the average transition time $\tau_K$ between minima in the unperturbed system~\cite{stochastic_resonance_review}. (Instead, the frequency-dependent response at fixed $T$ would only show a broad increase  for frequencies below approximately the transition rate $1/\tau_K$). At $T=T_{\rm sr}$ it is most probable for the ion to jump synchronously with the driving; for lower noise intensities the fluctuations are too weak to cause a transition, whereas for higher intensities transitions  will happen too frequently and thus more incoherently.

\begin{figure}
\includegraphics[width=\columnwidth]{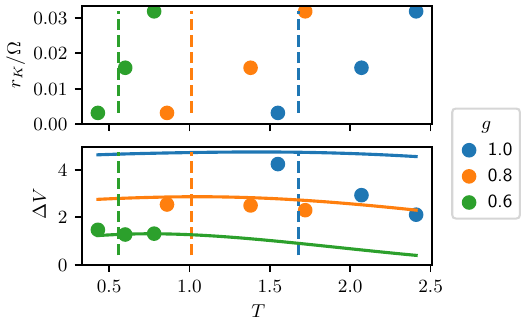}
\caption{\label{fig3} \justifying  
(a) Estimate of the transition rate for different values of $g$, each one corresponding to a different color. For each set of curves in Figure~\ref{fig3} with given $\Omega_p$, the transition rate at the temperature of the peak is approximated as $r_K\approx\Omega_p/\pi$. (b) Effective potential barrier as a function of temperature. The circles are extracted from from the points in (a) by inverting the Kramers formula~\eqref{eq:kramers}. The solid lines are the values of the energy barrier predicted by the impurity model. In both panels, the vertical dashed lines indicate the critical temperatures of transition to the ordered system as predicted by the semiclassical approach (see Figure~\ref{fig1}a).}
\end{figure}

By scanning $T$ and $\Omega_p$ one can thus measure the rate $\tau_K(T)$, as shown in the upper panel of Fig.~\ref{fig3}. The transition rate $r_K=1/\tau_K\,(T)$ even at $T=T_c$ remains well separated from the bare oscillator frequency $\Omega$, with $\pi r_K\,(T_c)\approx 0.05 \Omega$. Moreover,  $r_K\,(T)$ smoothly evolves through $T=T_c$.  This can be taken as a clear signature of a disordering transition, where the bistable potential persists in the disordered state.
(Note that in the present example the  stochastic resonance is probed by a spatially homogeneous probe, while the soft mode of the ordering is not at the Brillouin zone center. This would be different for a ferroelectric transition, which involve softening of a mode at the zone center.)

Following the measurement of the stochastic tunnelling time, one can therefore 
ask whether even the potential barrier itself can be determined from the data.  For an overdamped particle in an ideal quartic double well potential, the transition rate $r_K$ is given by the exact expression due to Kramers~\cite{kramers,kramers_review}, 
\begin{equation}
\label{eq:kramers}
r_K = \frac{\omega_0\omega_m}{2\pi\Omega\Gamma}\exp{\Bigl(-\frac{\Delta V}{k_B T}\Bigr)},
\end{equation}
where $\omega_0$ and $\omega_m$ are the frequencies of oscillation respectively in the center and in the minima of the potential, and $\Delta V$ is the energy barrier. While the experiment would determine  $r_K$, an independent quantitative measurement of $\Delta V$ therefore requires knowledge of $\Gamma$
 and the shape of the potential. Neverthless, assuming that the latter depends weakly on $T$, one can  extract a trend in $T$-dependence of $\Delta V$. In the lower panel of Fig.~\ref{fig3} an estimate of $\Delta V$ is reported, obtained by replacing $\omega_0=\omega_m=\Omega$ in \eqref{eq:kramers}. The result again confirms the existence of a potential $\Delta V$ that remains intact through $T_c$, with a slight decrease in $T$. The order of magnitude of this estimate is also  consistent with the barrier in the mean field potential (solid lines in the same Figure).

{\em Conclusion --}
In conclusion, we have shown that the signature of stochastic resonance can serve as a unique probe of emergent local multi-stability in inhomogeneous phases, which occur as precursors to symmetry breaking or as transient states following ultrafast disruption of long-range order. Although stochastic resonance is well known in statistical and non-linear physics, the setting presented here is distinct because (i) both the noise and the bistable potential arise intrinsically from electron-lattice interactions rather than external sources, and (ii) an effective coupling between lattice oscillators is mediated by the electrons themselves. Despite these differences, a clear stochastic resonance signal emerges, as a peak in the linear response plotted against temperature. This enables a direct measurement of the average tunneling time between local minima and, indirectly, shows the existence of a potential barrier even in the symmetric phase. The present work is specifically based on the Holstein model in the DMFT limit ($Z \to \infty$), but the results suggest exploring the use of SR as a probe of more complex systems, such as the proposed inhomogeneous disordered phases in VO$_2$~\cite{wall:ultrafast_disordering_VO2-science,wall:ultrafast_xray_V02-nature_physics,wall:ultrafast_disordering_collisional_forces-nature_physics}, V$_2$O$_3$~\cite{wall:prl} and manganites~\cite{wall:disorder_C4-nature_comm} (these may involve more than two discrete states).  On the technical side, interesting extensions would be to map the stochastic DMFT on a Fokker-Planck equation \cite{gang:fokker_planck_stoc_res}, to link to molecular dynamics simulations of CDW systems~\cite{rossi_wehling:molecular_dynamics,freericks:pump_probe_cdw}, and to extend the stochastic dynamics to strongly correlated (Mott) systems.

\acknowledgements
We thank Igor Krivenko for helpful discussion about the DMFT code, and Michael Fechner for insightful exchange about stochastic resonance. F.V. and M.E. are supported by the Cluster of Excellence ``CUI: Advanced Imaging of Matter'' of the Deutsche Forschungsgemeinschaft (DFG) -- EXC 2056 -- Project No. 390715994. A.P. acknowledges funding from the European Union's Horizon 2020 research and innovation programme under the Marie Sklodowska-Curie Postdoctoral Fellowship (Grant Agreement No. 101149691 - DISRUPT).

\clearpage
\appendix
\widetext
\begin{center}
\textbf{\large \centering Supplemental Material: Stochastic resonance in disordered charge-density-wave systems}
\end{center}
\vspace{0.0cm}
\setcounter{equation}{0}
\setcounter{figure}{0}
\setcounter{table}{0}
\setcounter{page}{1}

\section{Impurity model mean-field potential}
\label{sec:imp_potential}
The mean field description of the impurity model is analogous to a \emph{frozen-phonon} picture, where the ion position becomes a classical quantity acting on the electrons as an external potential \cite{picano:semiclassical_theory}. The retarded electronic Green's function becomes
\begin{equation}
    G^R_X\,(\omega) = \bigl(\omega - \sqrt{2}gX-\Delta^R\,(\omega)+i0^+\bigr)^{-1}\,,
\end{equation}
where the hybridization function of the bath was also introduced as having a semicircular density of states of width $W$:
\begin{equation}
\label{eq:hyb_semicirc}
\Delta^R\,(\omega) = \alpha\times
\begin{cases}
    \omega - \sqrt{\omega^2 - \frac{W^2}{4}}&\textup{if $\frac{W}{2}<\omega$}\,; \\
    \omega - i\sqrt{\frac{W^2}{4}-\omega^2}&\textup{if $-\frac{W}{2}<\omega\le\frac{W}{2}$}\,;\\
    \omega + \sqrt{\omega^2 - \frac{W^2}{4}}&\textup{if $\omega\le-\frac{W}{2}$}\,.
\end{cases}
\end{equation}
with
\begin{equation}
    \alpha = \frac{8\lvert J_0\rvert^2}{W^2}\,.
\end{equation}
The expression~\ref{eq:hyb_semicirc} can be obtained by assuming a semicircular bath spectral function 
\begin{align}
    \label{eq:spectral_function_bath}
    A_\Delta\,(\omega) = -\frac{1}{\pi}\Im{\bigl(\Delta\,(\omega)\bigr)}
\end{align}
and calculating the real part through Kramers-Kronig relations. The impurity spectral function can be separeted into a continuous and singular part
\begin{gather}
    \label{eq:spectral_function}
    A_X\,(\omega) = -\frac{1}{\pi}\Im \,\bigl(G_X\,(\omega)\bigr) = A_X^{\textup{(r)}}\,(\omega) + A_X^{\textup{(s)}}\,(\omega)
\end{gather}
where
\begin{align}
    A_X^{\textup{(r)}}\,(\omega) &= \frac{A_\Delta\,(\omega)}{(\omega-\sqrt{2}gX)^2 + \pi^2 A_\Delta^2\,(\omega)} \,; \\
    A_X^{\textup{(s)}}\,(\omega) &= Z_s\delta\,(\omega-\omega_s)\,.
\end{align}
The position of the singularity is determined by solving the equation
\begin{equation}
    \omega_s -\sqrt{2}gX -\Re\,\bigl(\Delta^R\,(\omega_s)\bigr) = 0\,,
\end{equation}
with the additional condition $\lvert\omega_s\rvert>W/2$, and the  weight of the singular part is
\begin{equation}
    Z_s = \biggl( 1 - \alpha \times\Bigl( 1 - \frac{\lvert \omega_s\rvert}{\sqrt{\omega_s ^2-W^2/4}} \Bigr)\biggr)^{-1}\,.
\end{equation}
The potential is initially a sum of four terms:
\begin{equation}
    V_\textup{imp}\,(X) = E_\textup{ph}\,(X) + E_\textup{kin}\,(X) + E_\textup{int}\,(X) +
    E_\textup{on-site}\,(X)\,,
    \nonumber
\end{equation}
where 
\begin{align}
    E_\textup{ph}\,(X) = \frac{1}{2}\Omega X^2 \,,
    \end{align}
    is the bare phonon potential, and 
\begin{align}    
         E_\textup{kin}\,(X) &= 4\int_{-\infty}^\infty\mathrm{d}\omega\Bigl(\Re\,\bigl(\Delta^R\,(\omega)\bigr)A_X\,(\omega) +\Re\,\bigl(G^R\,(\omega)\bigr)A_\Delta\,(\omega)\Bigr)f_\beta\,(\omega)\,,
         \\
    E_\textup{int}\,(X) &= \sqrt{2} gX\Bigl(2\int_{-\infty}^{\infty}\!\!\!\!\mathrm{d}\omega A_X\,(\omega)f_\beta\,(\omega) - 1\Bigr) \,,\\ 
    E_\textup{on-site}\,(X) &= -2\mu\int_{-\infty}^{\infty}\mathrm{d}\omega A_X\,(\omega)f_\beta\,(\omega)\,.
\end{align}
The function $f_\beta$ is the Fermi distribution at the inverse temperature $\beta$ and zero chemical potential.

Substituting the expression for the hybridization and impurity spectral function yields
\begin{align}
V_\textup{imp}\,(X) & = \frac{1}{2}\Omega^2X^2+4I_1\,(X) + 2(\mu-\sqrt{2}gX)I_2\,(X)- \sqrt{2} g X + E_\textup{s}\,(X),\,
\label{eq:V_imp_final_expression}
\end{align}
with the moments 
\begin{gather}
    I_1\,(X) = \int_{-W/2}^{W/2} \mathrm{d}\omega f_\beta\,(\omega) A_X^{(r)}\,(\omega)\omega\,, \\
    I_2\,(X) = \int_{-W/2}^{W/2} \mathrm{d}\omega f_\beta\,(\omega) A_X^{(r)}\,(\omega),\,.
\end{gather}
and a contribution coming from the singular part of the spectrum,
\begin{align}
    E_\textup{s}\,(X) = Z_s f_\beta\,(\omega_s)&\biggl( 4\alpha\times\Bigl(\omega_s-\frac{\omega_s}{\lvert\omega_s\rvert}\sqrt{\omega_s - \frac{W^2}{4}}\Bigr)- 2(\mu-\sqrt{2}g X)\biggr)\,.
\end{align}
The expression~\eqref{eq:V_imp_final_expression} can then be used to plot the potential curves and the Boltzmann distributions in Figure~1b and~1c of the main text.

\section{Electronic density in the adiabatic approximation}
\label{sec:el_dens_adiab}
As stated in the main text, the electronic density appearing in the expression of the electronic force~(3) was calculated within an adiabatic approximation. At every time $t$, the configuration $X=\{X_j\,(t)\}$ is given, and the electrons are assumed in equilibrium with respect to the resulting spatially varying onsite 
potential
\begin{equation}
    \epsilon_j\,(t) = \sqrt{2}gX_j\,(t)\,.
\end{equation}
The imaginary-time Matsubara Green's function is
\begin{equation}
\label{eq:matsubara_components}
G^M_{j}\,(i\omega_n) = (i\omega_n - \epsilon_j\,(t) - \Delta^M\,(i\omega_n;X) + i0^+)^{-1}
\end{equation}
with $\Delta$ the self-consistent hybridization~\eqref{eq:dmft2}. Substituting~\eqref{eq:matsubara_components} in the latter, leads to the closed equation
\begin{equation}
    \Delta^M\,(i\omega_n;X) = \frac{1}{N_\textup{ens}}\sum_j \frac{1}{i\omega_n - \epsilon_j\,(t) - \Delta^M\,(i\omega_n;X) }
\end{equation}
which can be solved at every time $t$ by iteration.

Once $\Delta$ is known, it can be substituted in~\eqref{eq:matsubara_components}. The electron density can then be computed with the relation
\begin{equation}
\label{eq:density_matsubara}
n_j\,(X) = G_j^M\,(\tau=0^-)\,.
\end{equation}
In order to compute $G_j^M\,(\tau=0^-)$ from the transform
\begin{equation}
\label{eq:matsubara}
G^M_{j}\,(\tau) = \frac{1}{\beta}\sum_n G^M_{j}\,(i\omega_n) e^{-i\omega_n\tau},
\end{equation}
it is convenient to rewrite
\begin{equation}
\label{eq:matsubara_component_approx}
    G^M_j\,(i\omega_n) = \frac{1}{i\omega_n} + \Bigl(G^M_j\,(i\omega_n) - \frac{1}{i\omega_n}\Bigr)\,.
\end{equation}
Substituting the latter expression in~\eqref{eq:matsubara}, the sum over the first term gives the electron density in absence of external field and of the bath, namely $1/2$. Since the result in~\eqref{eq:density_matsubara} is a real quantity, instead, for the sum of the terms in parentheses only the real part can be retained. In the end one retrieves
\begin{equation}
\label{eq:density_final}
    n_j\,(X) = \frac{1}{2} - \sum_n\frac{\epsilon_j+\Delta'\,(i\omega_n)}{\bigl(\Delta''\,(i\omega_n)-\omega_n\bigr)^2+\bigl(\epsilon_j+\Delta'\,(i\omega_n)\bigr)^2}\,,
\end{equation}
where $\Delta'$ and $\Delta''$ indicate respectively the real and imaginary part of $\Delta^M$ (the $X$-dependence was also omitted). The expression~\eqref{eq:density_final} is the one used in order to calculate the electronic force $F_\textup{el}$ defined in~(3) in the main text.

\section{Damping and diffusion coefficients}
\label{sec:damp_and_diff}
The damping and diffusion coefficients are calculated from the impurity model in the mean-field approximation, with a time-independent bath. As stated in the main text, they depend on the density-density correlation function in the presence of the $X$-field:
\begin{equation}
    \Pi_{j,X}\,(t_1,t_2) = -i\bigl\langle n_j(t_1)n_j(t_2)\rangle_X^\textup{conn} \,,
\end{equation}
where the times $t_1$ and $t_2$ are points on the Keldysh contour~\cite{Kamenev}. After a Keldysh rotation is performed, one can introduce the retarded and Keldsyh components and their Wigner transforms:
\begin{equation}
    \Pi^{K/R}_{j,X}\,(t,\omega) = \int_{-\infty}^\infty\!\mathrm{d}s \,e^{i\omega s}\Pi^{K/R}_{j,X}\,(t + s/2,t-s/2)\,.
\end{equation}
The formulas for the coefficients in the most general setting are:
\begin{gather}
\label{eq:damping}
    \Gamma_j\,(X,t) = -2g^2\partial_\omega\Bigl(\Im{\bigl(\Pi^R_{j,X}\,(t,\omega\bigr)}\Bigr)_{\omega=0}\,, \\
\label{eq:diffusion}
    K_j\,(X,t) = -g^2\Im{\bigl(\Pi^K_{j,X}\,(t,\omega)\bigr)}_{\omega=0}\,,
\end{gather}
as also derived in~\cite{picano:semiclassical_theory}.

Within the adiabatic approximation the correlation functions are calculated in equilibrium, therefore they become stationary (they depend only on the time-difference $t-t'$).
Moreover, as a further approximation, the lattice index $j$ is also removed and the coefficients are assumed to be the same for all lattice points; they are then calculated from the effective impurity model introduced in the main text and in the first Section of the Supplemental Material. The set of ion coordinates is then again reduced to a single $X$.

In equilibrium, the fluctuation-dissipation theorem implies
\begin{equation}
\label{eq:fdr_susceptibility}
    \Pi_X^K\,(\omega) = 2i\coth{\Bigl(\frac{1}{2}\beta\omega\Bigr)}\Im{\bigl(\Pi^R_X\,(\omega)\bigr)}\,.
\end{equation}
Moreover, since the system is non-interacting, the retarded component of the correlation function in real-time can be expressed as the product of Green's function
\begin{align}
    \Pi^R_X\,(t)    &= -i \bigl(G^R_X\,(t)G^K_X\,(-t) + G^K_X\,(t)G^A_X\,(-t)\bigr).
\end{align}
where $G^A(t)=G^R(-t)^\dagger$.
 Using then the convolution theorem to calculate the Fourier transform yields
\begin{align}
    \Im\bigl(\Pi^R_X\,(\omega)\bigr) =& \pi\int_{-\infty}^{\infty}\mathrm{d}\omega' \tanh{\Bigl(\frac{1}{2}\beta\omega'}\Bigr)A_X\,(\omega')\bigl(A_X\,(\omega'+\omega)-A_X\,(\omega'-\omega)\bigr)\,
\end{align}
where the fluctuation-dissipation relation for the Green's function was also exploited, namely
\begin{equation}
    G^K_X\,(\omega) = 2i\tanh{\Bigl(\frac{1}{2}\beta\omega\Bigr)}A_X\,(\omega)\,.
\end{equation}
Furthermore, the spectral function is defined as in~\eqref{eq:spectral_function}.

When the partial derivative with respect to $\omega$ is taken and then evaluated at $\omega=0$, the singular part of the spectral function does not contribute. The damping coefficient then reads as follows:
\begin{equation}
\label{eq:diffusion_calculation}
    \begin{split}
        \Gamma\,(X) &= -4\pi g^2 \int_{-\infty}^{\infty}\mathrm{d}\omega \tanh{\Bigl(\frac{1}{2}\beta\omega}\Bigr)A_X^{(r)}\,(\omega)(\partial_\omega A_X^{(r)})\,(\omega) \\
        &= -2\pi g^2\int_{-\infty}^{\infty}\mathrm{d}\omega \tanh{\Bigl(\frac{1}{2}\beta\omega}\Bigr)(\partial_\omega (A_X^{(r)})^2)\,(\omega) \\
        &=\pi\beta g^2\int_{-\infty}^{\infty} \mathrm{d}\omega\Bigl(1-\tanh^2{\Bigl(\frac{1}{2}\beta\omega}\Bigr)\Bigr)(A_X^{(r)})^2\,(\omega)\,,
    \end{split}
\end{equation}
where in the last row integration by part was used, assuming that the spectral function vanishes at infinity.

For the evaluation of the diffusion coefficient in~\eqref{eq:diffusion}, the fluctuation-dissipation relation~\eqref{eq:fdr_susceptibility} can be used. Since the hyperbolic cotangent diverges at the origin, the limit $\omega\to 0$ must be taken. Using L'Hopital rule to calculate it and then comparing the expression with~\eqref{eq:diffusion_calculation} leads to
\begin{equation}
\label{eq:Einstein_relation}
K\,(X) = 2\frac{\Gamma\,(X)}{\beta}\,,
\end{equation}
that is, the usual Einstein relation between the damping and the diffusion coefficients.

\end{document}